# Thickness-shear Vibration Frequencies of an Infinite Plate with a Generalized Material Property Grading along the Thickness

Ji Wang*, Wenliang Zhang, Dejin Huang, Tingfeng Ma, Jianke Du, Lijun Yi
Piezoelectric Device Laboratory, School of Mechanical Engineering & Mechanics, Ningbo University,
818 Fenghua Road, Ningbo, Zhejiang 315211, CHINA
*E-mail: wangji@nbu.edu.cn

*Abstract*—For quartz crystal resonators of thickness-shear type, the vibration frequency and mode shapes, which are key features of resonators in circuit applications, reflect the basic material and structural properties of the quartz plate and its variation with time under various factors such as erosive gases and liquids that can cause surface and internal damages and degradation of crystal blanks. The accumulated effects eventually will change the surface conditions in terms of elastic constants and stiffness and more importantly, the gradient of such properties along the thickness. This is a typical functionally graded materials (FGM) structure and has been studied extensively for structural applications under multiple loadings such as thermal and electromagnetic fields in recent years. For acoustic wave resonators, such studies are equally important and the wave propagation in FGM structures can be used in the evaluation and assessment of performance, reliability, and life of sensors based on acoustic waves such as the quartz crystal microbalances (QCM). Now we studied the thickness-shear vibrations of FGM plates with properties of AT-cut quartz crystal varying along the thickness in a general pattern represented by a trigonometric function with both sine and cosine functions of the thickness coordinate. The solutions are obtained by using Fourier expansion of the plate deformation. We also obtained the frequency changes of the fundamental and overtone modes which are strongly coupled for the evaluation of resonator structures with property variation or design to take advantages of FGM in novel applications.

## I. Introduction

In the analysis and design of quartz crystal resonators, it has been known that the material is homogenous and uniform to simplify the design process. In reality, as one way to improve the performance of quartz crystal resonators, variation of material configuration has been adopted to change the properties of vibration modes to satisfy needs of resonator performance. The variation of quartz crystal plate configuration presents challenges to the analysis and fabrication of quartz crystal resonators due to complicated material properties, but the needs are justified from the significant improvement of device performances and efforts have been made through design and fabrication. It is known that the variation will reduce the couplings of typical vibration modes such as the thickness-shear (TSh) and flexure which appear in the functioning vibrations of quartz crystal resonators. The difficulties for accurate analysis of anisotropic plate vibrations are generally known with complications from anisotropic material, coupled modes, and accompanying boundary conditions. These complications have been there and many techniques primarily based on various simplifications and approximations have been existed for solutions which are required in resonator design. Further refinement and improvement of analysis are made with the consideration of more physical complications like irregular configurations of crystal blanks such as the commonly known beveling, which refers to the contour or thickness variations in edges of blanks to suppress strong couplings between thickness-shear and flexural modes. It has been found that the beveling in common configurations such as linear or quadratic variation of thickness can weaken couplings between thickness-shear and flexural modes significantly [1-2]. As a result, processing techniques based on grinding have been developed to take the advantage by making contour or beveling in edges of crystal blanks, but not necessarily in any specific form of design [3]. Since material grading is equivalent to thickness variation, we started the analysis of vibration frequencies of an FGM quartz crystal plate in the thickness-shear modes to explore possible applications in sensor technology [4].

There have been studies on wave propagations in FGM solids with various methods including approximations and discrete techniques [4-6]. There is no doubt that such analysis is important in establishing the correlation between performance changes and FGM patterns in wave propagation and high frequency vibrations of device structures. In the material processing part, FGM patterns can be realized through modern technologies such as radiation, laser, and chemical etching, and so on. We can make the partial materials to FGM materials so advantages can be taken and the equivalent effect of beveling can be achieved with minimal cost in fabrication. In this case, we have finally utilized the FGM for possible advantages rather than known applications in structural protection and enhancement [6]. Of course, applications of FGM in acoustic wave devices have been studied before for possible performance enhancement and novel fabrication techniques [7-9]. The essential nature of acoustic wave devices requires the analysis to consider wave propagations or high frequency vibrations in piezoelectric materials and solids as have been done for FGM plates and structures [7-16]. Such studies have been carried out before and there are positive leads to be followed up with more research for effective FGM to meet performance requirements of next generation acoustic wave devices. In this paper, we make further assumption on the material grading to represent a general pattern which is no longer symmetric regarding the middle plane, or the thickness. By following the earlier approach based on the Fourier series expansion of the thickness-shear displacement which now has both symmetric and antisymmetric components, a more general example of vibrations of an infinite FGM plate is studied. The frequency equation of the FGM plate in the thickness-shear vibration mode is obtained for studies on effects of patterns and parameters of material grading.

## II. PROBLEM AND FORMULATION

We start with an infinite plate of quartz crystal with modified material properties resembling to FGM. The plate is shown in Fig.1, and the thickness of plate is $2b$.

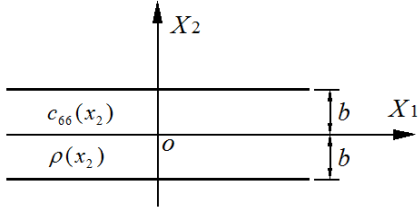

Fig. 1 An infinite quartz crystal plate

For a simplest problem of thickness-shear waves propagating in the elastic plate, the stress equation of motion is

$$(c_{66}u_{1,2})_{,2} = \rho(x_2)\ddot{u}_1, \qquad (1)$$

where $c_{66}$, $\rho$, and $u$ are elastic constant, density, and displacement of plate, respectively.

In case of $c_{66}$ and $\rho$ are constants, the simple solution is known as

$$u = \begin{cases} A \sin\frac{r\pi x_2}{2b} e^{i\omega t}, & r \text{ odd}, \\ B \cos\frac{r\pi x_2}{2b} e^{i\omega t}, & r \text{ even}, \end{cases} \qquad (2)$$

$$\omega = \frac{n\pi}{2b}\sqrt{\frac{c_{66}}{\rho}}, \quad n = 1, 2, \cdots, \qquad (3)$$

where $A(B)$ and $\omega$ are amplitudes and vibration frequency, respectively.

In this study, we assume that quartz crystal plate has been treated and constants $c_{66}$ and $\rho$ are now functions of the thickness coordinate $x_2$ in the form of

$$(c_{66}, \rho) = (C_{66}, \rho_0)f(x_2), f(x_2) = \alpha + \beta_1 \cos\frac{x_2}{Nb} + \beta_2 \sin\frac{x_2}{Nb}, \qquad (5)$$

where constants $C_{66}$, $\rho_0$, $\alpha$, $\beta_1$, and $\beta_2$ are known. Apparently, if integer $N$ is large enough, we get material properties approach to uniform. With appropriate integer $N$, we can study the effect of material property grading on the vibration frequency.

With the consideration of traction-free surfaces of the plate in Fig. 1, we can assume the thickness displacement as

$$u = \left(\sum_{n=1,3,5,\cdots}^{\infty} A_n \sin\frac{n\pi}{2b}x_2 + \sum_{m=0,2,4,\cdots}^{\infty} B_m \cos\frac{m\pi}{2b}x_2\right)e^{i\omega t}, \qquad (6)$$

Then through the expansion of (1) we have

$$c_{66,2}u_{,2} + c_{66}u_{,22} = \rho(x_2)\ddot{u}, \qquad (7)$$

and by substituting (6) into (7) we obtain

$$C_{66}\left(-\frac{\beta_1}{Nb}\sin\frac{x_2}{Nb} + \frac{\beta_2}{Nb}\cos\frac{x_2}{Nb}\right)\sum_{n=1,3,5,\cdots}^{\infty}\frac{n\pi}{2b}A_n\cos\frac{n\pi}{2b}x_2$$

$$-C_{66}\left(-\frac{\beta_1}{Nb}\sin\frac{x_2}{Nb} + \frac{\beta_2}{Nb}\cos\frac{x_2}{Nb}\right)\sum_{m=0,2,4,\cdots}^{\infty}\frac{m\pi}{2b}B_m\sin\frac{m\pi}{2b}x_2$$

$$-C_{66}\left(\alpha + \beta_1\cos\frac{x_2}{Nb} + \beta_2\sin\frac{x_2}{Nb}\right)\sum_{n=1,3,5,\cdots}^{\infty}\left(\frac{n\pi}{2b}\right)^2 A_n\sin\frac{n\pi}{2b}x_2$$

$$+\rho_0\omega^2\left(\alpha + \beta_1\cos\frac{x_2}{Nb} + \beta_2\sin\frac{x_2}{Nb}\right)\sum_{n=1,3,5,\cdots}^{\infty}A_n\sin\frac{n\pi}{2b}x_2$$

$$-C_{66}\left(\alpha + \beta_1\cos\frac{x_2}{Nb} + \beta_2\sin\frac{x_2}{Nb}\right)\sum_{m=0,2,4,\cdots}^{\infty}\left(\frac{m\pi}{2b}\right)^2 B_m\cos\frac{m\pi}{2b}x_2$$

$$+\rho_0\omega^2\left(\alpha + \beta_1\cos\frac{x_2}{Nb} + \beta_2\sin\frac{x_2}{Nb}\right)\sum_{m=0,2,4,\cdots}^{\infty}B_m\cos\frac{m\pi}{2b}x_2 = 0, \qquad (8)$$

With $f(x_2) = \alpha + \beta_1 \cos\frac{x_2}{Nb} + \beta_2 \sin\frac{x_2}{Nb}$, we also use

$$f'(x_2)\cos\frac{n\pi}{2b}x_2 = f(x_2)\frac{f'(x_2)}{f(x_2)}\cos\frac{n\pi}{2b}x_2, \qquad (9)$$

$$f'(x_2)\sin\frac{m\pi}{2b}x_2 = f(x_2)\frac{f'(x_2)}{f(x_2)}\sin\frac{m\pi}{2b}x_2. \qquad (10)$$

This actually is to obtain the Fourier expansion of functions in the form of

$$\frac{f'(x_2)}{f(x_2)}\cos\frac{n\pi}{2b}x_2 \sim \sum_{p=1,3,5,\cdots}^{\infty} A'_{np}\sin\frac{p\pi}{2b}x_2 + \sum_{q=0,2,4,\cdots}^{\infty} B'_{nq}\cos\frac{q\pi}{2b}x_2,$$

$$\frac{f'(x_2)}{f(x_2)}\sin\frac{m\pi}{2b}x_2 \sim \sum_{r=1,3,5,\cdots}^{\infty} A^*_{mr}\sin\frac{r\pi}{2b}x_2 + \sum_{s=0,2,4,\cdots}^{\infty} B^*_{ms}\cos\frac{s\pi}{2b}x_2,$$

$$(11)$$

where $A'_{np}, B'_{nq}, A^*_{mr}, B^*_{ms}$ are coefficients in the form of

$$A'_{np} = \frac{1}{b}\int_{-b}^{b}\frac{f'(x_2)}{f(x_2)}\cos\frac{n\pi}{2b}x_2 \sin\frac{p\pi}{2b}x_2\, dx_2, p = 1,3,5,\cdots,$$

$$B'_{n0} = \frac{1}{2b}\int_{-b}^{b}\frac{f'(x_2)}{f(x_2)}\cos\frac{n\pi}{2b}x_2\, dx_2,$$

$$B'_{nq} = \frac{1}{b}\int_{-b}^{b}\frac{f'(x_2)}{f(x_2)}\cos\frac{n\pi}{2b}x_2 \cos\frac{q\pi}{2b}x_2\, dx_2, q = 2,4,\cdots, \quad (12)$$

$$A^*_{mr} = \frac{1}{b}\int_{-b}^{b}\frac{f'(x_2)}{f(x_2)}\sin\frac{m\pi}{2b}x_2 \sin\frac{r\pi}{2b}x_2\, dx_2, r = 1,3,5,\cdots,$$

$$B^*_{m0} = \frac{1}{2b}\int_{-b}^{b}\frac{f'(x_2)}{f(x_2)}\sin\frac{m\pi}{2b}x_2\, dx_2,$$

$$B^*_{ms} = \frac{1}{b}\int_{-b}^{b}\frac{f'(x_2)}{f(x_2)}\sin\frac{m\pi}{2b}x_2 \cos\frac{s\pi}{2b}x_2\, dx_2, s = 2,4,\cdots.$$

Clearly,

$$A'_{np} = B^*_{pn}. \quad (13)$$

Different notations are used for the convenience of writing and programming.

By substituting (11), (12) into (8), we have

$$\sum_{n=1,3,5,\cdots}^{\infty} f_n(A,B)\sin\frac{n\pi x_2}{2b} + \sum_{m=0,2,4,\cdots}^{\infty} g_m(A,B)\cos\frac{m\pi x_2}{2b} = 0, \quad (14)$$

where

$$g_0(A,B) = -\frac{\pi C_{66}}{2b}(B'_{10}A_1 + 3B'_{30}A_3 + \cdots + nB'_{n0}A_n) + \frac{\pi C_{66}}{2b}(2B^*_{20}B_2 + 4B^*_{40}B_4 + \cdots + mB^*_{m0}B_m) - \rho_0\omega^2 B_0,$$

$$f_1(A,B) = -\frac{\pi C_{66}}{2b}(A'_{11}A_1 + 3A'_{31}A_3 + \cdots + nA'_{n1}A_n) + \frac{\pi C_{66}}{2b}(2A^*_{21}B_2 + 4A^*_{41}B_4 + \cdots + mA^*_{m1}B_m) + \left[\left(\frac{\pi}{2b}\right)^2 C_{66} - \rho_0\omega^2\right]A_1,$$

$$g_2(A,B) = -\frac{\pi C_{66}}{2b}(B'_{12}A_1 + 3B'_{32}A_3 + \cdots + nB'_{n2}A_n) + \frac{\pi C_{66}}{2b}(2B^*_{22}B_2 + 4B^*_{42}B_4 + \cdots + mB^*_{m2}B_m) + \left[2^2 C_{66}\left(\frac{\pi}{2b}\right)^2 - \rho_0\omega^2\right]B_2,$$

$$f_3(A,B) = -\frac{\pi C_{66}}{2b}(A'_{13}A_1 + 3A'_{33}A_3 + \cdots + nA'_{n3}A_n) + \frac{\pi C_{66}}{2b}(2A^*_{23}B_2 + 4A^*_{43}B_4 + \cdots + mA^*_{m3}B_m) + \left[3^2\left(\frac{\pi}{2b}\right)^2 C_{66} - \rho_0\omega^2\right]A_3,$$

$$g_4(A,B) = -\frac{\pi C_{66}}{2b}(B'_{14}A_1 + 3B'_{34}A_3 + \cdots + nB'_{n4}A_n) + \frac{\pi C_{66}}{2b}(2B^*_{24}B_2 + 4B^*_{44}B_4 + \cdots + mB^*_{m4}B_m) + \left[4^2 C_{66}\left(\frac{\pi}{2b}\right)^2 - \rho_0\omega^2\right]B_4,$$

$$\vdots$$

$$f_n(A,B) = -\frac{\pi C_{66}}{2b}(A'_{1n}A_1 + 3A'_{3n}A_3 + \cdots + nA'_{nn}A_n) + \frac{\pi C_{66}}{2b}(2A^*_{2n}B_1 + 4A^*_{4n}B_2 + \cdots + mA^*_{mn}B_m) + \left[n^2\left(\frac{\pi}{2b}\right)^2 C_{66} - \rho_0\omega^2\right]A_n,$$

$$g_m(A,B) = -\frac{\pi C_{66}}{2b}(B'_{1m}A_1 + 3B'_{3m}A_3 + \cdots + nB'_{nm}A_n) + \frac{\pi C_{66}}{2b}(2B^*_{2m}B_2 + 4B^*_{4m}B_4 + \cdots + mB^*_{mm}B_m) + \left[m^2 C_{66}\left(\frac{\pi}{2b}\right)^2 - \rho_0\omega^2\right]B_m.$$

(15)

For (14) to be true, we must have the coefficients vanish separately, i.e.

$$f_n(A,B) = 0, g_m(A,B) = 0, n = 1,3,5,\cdots; m = 0,2,4,\cdots. \quad (16)$$

With the normalized parameters

$$\Omega = \frac{\omega}{\omega_1}, \Omega^2 = \frac{\omega^2}{\omega_1^2} = \rho_0\omega^2 \frac{4b^2}{\pi^2 C_{66}}, \quad (17)$$

where $\omega_1$ is the fundamental thickness-shear frequency of a plate with uniform properties given in (3).

Now we have a system of equations for amplitudes of displacements as

$$K \cdot S = 0, \quad (18)$$

where $K$ and $S$ are matrices as

$$K = \begin{bmatrix} -\Omega^2 & -\frac{2b}{\pi}B'_{10} & \frac{2b}{\pi}2B^*_{20} & \cdots & -\frac{2b}{\pi}nB'_{n0} & \frac{2b}{\pi}mB^*_{m0} \\ 0 & -\frac{2b}{\pi}A'_{11}+1-\Omega^2 & \frac{2b}{\pi}2A^*_{21} & \cdots & -\frac{2b}{\pi}nA'_{n1} & \frac{2b}{\pi}mA^*_{m1} \\ 0 & -\frac{2b}{\pi}B'_{12} & \frac{2b}{\pi}2B^*_{22}+2^2-\Omega^2 & \cdots & -\frac{2b}{\pi}nB'_{n2} & \frac{2b}{\pi}mB^*_{m2} \\ \vdots & \vdots & \vdots & \ddots & \vdots & \vdots \\ 0 & -\frac{2b}{\pi}A'_{1n} & \frac{2b}{\pi}2A^*_{2n} & \cdots & -\frac{2b}{\pi}nA'_{nn}+n^2-\Omega^2 & \frac{2b}{\pi}mA^*_{mn} \\ 0 & -\frac{2b}{\pi}B'_{1m} & \frac{2b}{\pi}2B^*_{2m} & \cdots & -\frac{2b}{\pi}nB'_{nm} & \frac{2b}{\pi}mB^*_{mm}+m^2-\Omega^2 \end{bmatrix},$$

(19)

$$S = \{B_0 \quad A_1 \quad B_2 \quad \cdots \quad A_n \quad B_m\}^T. \quad (20)$$

For free vibrations of an infinite FGM quartz crystal plate, frequencies can be obtained by setting the determinant of $K$ matrix to vanish through

$$|K| = 0. \quad (21)$$

With all known parameters of material properties, we can evaluate (21) for the fundamental and overtone vibration frequencies. It is clear from (19) that the fundamental and overtone modes are now closely coupled. In other words, the FGM plates can no longer support pure modes of thickness-shear vibrations even it is infinite. In this case, our interests will be on the effects of FGM patterns on frequencies of the coupled thickness-shear modes.

### III. NUMERICAL RESULTS

With the frequency equation in (21), we can calculate vibration frequencies of an FGM plate with given property grading pattern and parameters. Since the frequency equation cannot be given in an explicit form, we have to perform the calculation with estimated number of modes, or order of the matrix, to obtain convergent solutions with given limit for the evaluation of frequency. In the calculation, parameters $\alpha$, $\beta_1$ and $\beta_2$ will have effects on the frequency solution. With $\alpha = 1$ and $\beta_1$ and $\beta_2$ as zeroes or very small numbers, the plate can be considered as a uniform one. Parameter $N$ also has great effect on the frequency through its variation. It is clear that the smaller $N$ represents a smooth variation of material properties across the thickness, and very large $N$ implies a plate with very small variation of properties, or the plate can be considered an a uniform one.

As a numerical example, we set $\alpha = \beta_1 = \beta_2 = 1$, which means a typical FGM plate with relatively large variations of material properties. We made the calculations with different parameter $N$ to demonstrate effects of grading and the solution procedure. It is also shown the test procedure with estimated numbers of modes to get accurate solutions.

TABLE I. VIBRATION FREQUENCIES WITH THE ORDER OF DETERMINANT = 3

| Mode no. | $N$=5 | $N$=10 | $N$=15 | $N$=20 | $N$=25 |
|---|---|---|---|---|---|
| 1 | 1.003534 | 1.000882 | 1.000392 | 1.00022 | 1.000141 |
| 2 | 2.001288 | 2.00032 | 2.000142 | 2.00008 | 2.000051 |

TABLE II. VIBRATION FREQUENCIES WITH THE ORDER OF DETERMINANT = 5

| Mode no. | $N$=5 | $N$=10 | $N$=15 | $N$=20 | $N$=25 |
|---|---|---|---|---|---|
| 1 | 1.003555 | 1.000886 | 1.000394 | 1.000221 | 1.000142 |
| 2 | 2.001769 | 2.000439 | 2.000195 | 2.000109 | 2.00007 |
| 3 | 3.001167 | 3.00029 | 3.000129 | 3.000072 | 3.000046 |
| 4 | 4.000413 | 4.000103 | 4.000046 | 4.000026 | 4.000016 |

TABLE III. VIBRATION FREQUENCIES WITH THE ORDER OF DETERMINANT = 7

| Mode no. | $N$=5 | $N$=10 | $N$=15 | $N$=20 | $N$=25 |
|---|---|---|---|---|---|
| 1 | 1.003559 | 1.000887 | 1.000394 | 1.000222 | 1.000142 |
| 2 | 2.001788 | 2.000443 | 2.000197 | 2.000111 | 2.000071 |
| 3 | 3.001191 | 3.000295 | 3.000131 | 3.000074 | 3.000047 |
| 4 | 4.000872 | 4.000216 | 4.000096 | 4.000054 | 4.000034 |
| 5 | 5.000688 | 5.000171 | 5.000076 | 5.000043 | 5.000027 |

TABLE IV. VIBRATION FREQUENCIES WITH THE ORDER OF DETERMINANT = 9

| Mode no. | $N$=5 | $N$=10 | $N$=15 | $N$=20 | $N$=25 |
|---|---|---|---|---|---|
| 1 | 1.00356 | 1.000887 | 1.000394 | 1.000222 | 1.000142 |
| 2 | 2.001792 | 2.000444 | 2.000197 | 2.000111 | 2.000071 |
| 3 | 3.001195 | 3.000296 | 3.000131 | 3.000074 | 3.000047 |
| 4 | 4.000892 | 4.000221 | 4.000098 | 4.000055 | 4.000035 |
| 5 | 5.000711 | 5.000176 | 5.000078 | 5.000044 | 5.000028 |

TABLE V. VIBRATION FREQUENCIES WITH THE ORDER OF DETERMINANT = 11

| Mode no. | $N$=5 | $N$=10 | $N$=15 | $N$=20 | $N$=25 |
|---|---|---|---|---|---|
| 1 | 1.00356 | 1.000887 | 1.000394 | 1.000222 | 1.000142 |
| 2 | 2.001793 | 2.000444 | 2.000197 | 2.000111 | 2.000071 |
| 3 | 3.001197 | 3.000296 | 3.000131 | 3.000074 | 3.000047 |
| 4 | 4.000896 | 4.000222 | 4.000098 | 4.000055 | 4.000035 |
| 5 | 5.000716 | 5.000177 | 5.000079 | 5.000044 | 5.000028 |

TABLE VI. VIBRATION FREQUENCIES WITH THE ORDER OF DETERMINANT = 13

| Mode no. | $N$=5 | $N$=10 | $N$=15 | $N$=20 | $N$=25 |
|---|---|---|---|---|---|
| 1 | 1.00356 | 1.000887 | 1.000394 | 1.000222 | 1.000142 |
| 2 | 2.001794 | 2.000444 | 2.000197 | 2.000111 | 2.000071 |
| 3 | 3.001197 | 3.000296 | 3.000131 | 3.000074 | 3.000047 |
| 4 | 4.000898 | 4.000222 | 4.000098 | 4.000055 | 4.000035 |
| 5 | 5.000718 | 5.000178 | 5.000079 | 5.000044 | 5.000028 |

TABLE VII. VIBRATION FREQUENCIES WITH THE ORDER OF DETERMINANT = 15

| Mode no. | $N$=5 | $N$=10 | $N$=15 | $N$=20 | $N$=25 |
|---|---|---|---|---|---|
| 1 | 1.003561 | 1.000887 | 1.000394 | 1.000222 | 1.000142 |
| 2 | 2.001794 | 2.000445 | 2.000197 | 2.000111 | 2.000071 |
| 3 | 3.001198 | 3.000296 | 3.000131 | 3.000074 | 3.000047 |
| 4 | 4.000898 | 4.000222 | 4.000099 | 4.000055 | 4.000035 |

| | 5 | 5.000718 | 5.000178 | 5.000079 | 5.000044 | 5.000028 |

TABLE VIII. VIBRATION FREQUENCIES WITH THE ORDER OF DETERMINANT = 17

| Mode no. | $N=5$ | $N=10$ | $N=15$ | $N=20$ | $N=25$ |
|---|---|---|---|---|---|
| 1 | 1.003561 | 1.000887 | 1.000394 | 1.000222 | 1.000142 |
| 2 | 2.001794 | 2.000445 | 2.000197 | 2.000111 | 2.000071 |
| 3 | 3.001198 | 3.000296 | 3.000132 | 3.000074 | 3.000047 |
| 4 | 4.000899 | 4.000222 | 4.000099 | 4.000055 | 4.000035 |
| 5 | 5.000719 | 5.000178 | 5.000079 | 5.000044 | 5.000028 |

TABLE IX. VIBRATION FREQUENCIES WITH THE ORDER OF DETERMINANT = 19

| Mode no. | $N=5$ | $N=10$ | $N=15$ | $N=20$ | $N=25$ |
|---|---|---|---|---|---|
| 1 | 1.003561 | 1.000887 | 1.000394 | 1.000222 | 1.000142 |
| 2 | 2.001795 | 2.000445 | 2.000197 | 2.000111 | 2.000071 |
| 3 | 3.001198 | 3.000296 | 3.000132 | 3.000074 | 3.000047 |
| 4 | 4.000899 | 4.000222 | 4.000099 | 4.000055 | 4.000035 |
| 5 | 5.000719 | 5.000178 | 5.000079 | 5.000044 | 5.000028 |

TABLE X. VIBRATION FREQUENCIES WITH THE ORDER OF DETERMINANT = 21

| Mode no. | $N=5$ | $N=10$ | $N=15$ | $N=20$ | $N=25$ |
|---|---|---|---|---|---|
| 1 | 1.003561 | 1.000887 | 1.000394 | 1.000222 | 1.000142 |
| 2 | 2.001795 | 2.000445 | 2.000197 | 2.000111 | 2.000071 |
| 3 | 3.001198 | 3.000296 | 3.000132 | 3.000074 | 3.000047 |
| 4 | 4.000899 | 4.000222 | 4.000099 | 4.000055 | 4.000035 |
| 5 | 5.000719 | 5.000178 | 5.000079 | 5.000044 | 5.000028 |

From tables above, we found that for specific $N=5, 10, 15, 20, 25$ tested, we can always obtain accurate solutions for the first five modes with $K$ of order 17. The results based on this calculation have been summarized in Table XI. Again, it is clear that the overtone frequencies are no longer multiples of the fundamental frequency. With $N$ takes a large value, we obtain the exact solutions of uniform plates.

TABLE XI. EFFECTS OF FGM GRADING ON PLATE VIBRATION FREQUENCIES

| FGM index | 1 | 2 | 3 | 4 | 5 |
|---|---|---|---|---|---|
| $N=5$ | 1.003561 | 2.001795 | 3.001198 | 4.000899 | 5.000719 |
| $N=10$ | 1.000887 | 2.000445 | 3.000296 | 4.000222 | 5.000178 |
| $N=15$ | 1.000394 | 2.000197 | 3.000132 | 4.000099 | 5.000079 |
| $N=20$ | 1.000222 | 2.000111 | 3.000074 | 4.000055 | 5.000044 |
| $N=25$ | 1.000142 | 2.000071 | 3.000047 | 4.000035 | 5.000028 |
| $N=\infty$ | 1.000000 | 2.000000 | 3.000000 | 4.000000 | 5.000000 |

## IV. CONCLUSIONS

As the continuation of our study on the thickness-shear vibrations of quartz crystal plates with material grading across the thickness, we extended our earlier results to more general variation of material properties represented by a trigonometric function with both symmetric and anti-symmetric elements included. The solutions are obtained by expanding the deformation into Fourier series and through forcing the coefficients of coupled terms to vanish in order to satisfy the equation of motion. Then, the frequencies from the coefficient matrix are obtained as the vibration frequencies of the thickness-shear modes including the fundamental one and the overtones. In numerical examples, only the asymmetric vibrations are obtained and the procedure of calculation is examined. The complications of vibration modes through the couplings of overtone modes are observed. Such results can be used in the analysis of frequency spectra of resonators with both surface and internal damages and corrosions to help the understanding of certain measurement results in applications like sensors for chemical and biological samples. Of course, it is also possible to create material variation patterns of quartz crystal resonators to enhance mode couplings so measurements can be performed with different frequencies for validations or enable multiple functions with one device.

ACKNOWLEDGMENT

This research is supported by the National Natural Science Foundation of China through grants 11372145 and 11372146.